\documentclass[12pt,fleqn]{article}
\usepackage{amssymb}
\usepackage{graphicx}
\usepackage{bigints}
\usepackage{times}
\usepackage{lineno}
\usepackage{hyperref}
\usepackage{movie15}
\usepackage[sort&compress,comma]{natbib}
\usepackage{setspace}
\begin{document}
\def\D{\Delta}
\def\d{\delta}
\def\r{\rho}
\def\p{\pi}
\def\a{\alpha}
\def\g{\gamma}
\def\ra{\rightarrow}
\def\s{\sigma}
\def\b{\beta}
\def\e{\epsilon}
\def\G{\Gamma}
\def\om{\omega}
\def\l{\lambda}
\def\f{\phi}
\def\w{\psi}
\def\m{\mu}
\def\t{\tau}
\def\c{\chi}
\title{}

\title{\sffamily\textbf{A note on the complexity of evolutionary dynamics in a classic consumer-resource model}}
\vspace{1.5cm}
\author{\vspace*{-0mm} \\
Iaroslav Ispolatov$^\ast$ \& Michael Doebeli$^{\ast\ast}$\\
\vspace*{1mm}{\normalsize $^{\ast}$ Departamento de Fisica, Universidad de Santiago de Chile}
\vspace*{-2mm}\\
{\normalsize Casilla 302, Correo 2, Santiago, Chile; jaros007$@$gmail.com}
\vspace*{-2mm}\\
{\normalsize $^{\ast\ast}$ Departments of Zoology and Mathematics, University
of British Columbia,}
\vspace*{-2mm}\\
{\normalsize 6270 University Boulevard, Vancouver B.C. Canada, V6T 1Z4; doebeli$@$zoology.ubc.ca}\\
}
\vskip 1cm
\date{\vspace{10mm}\normalsize\today}
\maketitle
\vskip 10mm
\noindent 
\vskip 10 mm
{\bf RH:} Consumer-resource dynamics
\noindent 
\vskip 10 mm
{\bf Corresponding author:} Michael Doebeli, Department of Zoology and Department of
  Mathematics,  University of British Columbia, 6270 University Boulevard, Vancouver B.C. Canada, V6T 1Z4, Email: doebeli@zoology.ubc.ca.\\

\newpage
\noindent
{\bf \large Abstract}
\vskip 0.3cm
\noindent  We study how the complexity of evolutionary
 dynamics in the classic MacArthur consumer-resource model depends on resource uptake and utilization rates. The traditional assumption in such models is that the utilization rate of the consumer is proportional to the uptake rate. More generally, we show that if these two rates are related through a power law (which includes the traditional assumption as a special case), then the resulting evolutionary dynamics in the consumer is necessarily a simple hill-climbing process leading to an evolutionary equilibrium, regardless of the dimension of phenotype space. When utilization and uptake rates are not related by a power law, more complex evolutionary trajectories can occur, including the chaotic dynamics observed in previous studies for high-dimensional phenotype spaces. These results draw attention to the importance of distinguishing between utilization and uptake rates in consumer-resource models. 


\maketitle


\newpage

\section{Introduction}

Adaptive dynamics \citep{metz_etal1992,dieckmann_law1996,geritz_etal1998} has emerged as a useful framework to study long-term evolution as a dynamical system in phenotype space. Many interesting evolutionary phenomena have been studied using this framework, such as adaptive diversification and speciation (e.g. \cite{dieckmann_doebeli1999, dieckmann_etal2004, doebeli2011}), the evolution of cooperation \citep{doebeli_etal2004}, evolutionary suicide \citep{gyllenberg_parvinen2001}, and predator-prey arms races \citep{dieckmann_etal1995,doebeli_dieckmann2000, dercole_etal2006}. The framework is in principle designed to study evolution in high-dimensional phenotype spaces \citep{leimar2009} , but this line of research has only been taken up rather recently \citep{doebeli_ispolatov2010, doebeli_ispolatov2014, svardal_etal2014, ito_dieckmann2014, ispolatov_etal2016, geritz_etal2016, regocosta_etal2017}. In \cite{doebeli_ispolatov2014}, we have used a classical logistic competition model to argue that evolutionary dynamics tend to be complicated and chaotic in high-dimensional phenotype spaces. The competition model was based on heuristic assumptions about the form of the carrying capacity as a functions of phenotype, as well as about the form of the competition kernel as a function of the phenotypes of competing individuals. It is known that even in its simplest form, this competition model cannot be derived from an underlying consumer-resource model \citep{ackermann_doebeli2004}. It is therefore interesting to see what kind of evolutionary dynamics occur when the dynamics of the resource that mediates competition is taken into account.

The MacArthur model \citep{macarthur_levins1967} is a classic consumer-resource model that has been used in numerous evolutionary studies (e.g. \cite{roughgarden1979, slatkin1980,case1981, ackermann_doebeli2004, svardal_etal2014}). It was initially used to study the ecological dynamics of competition for a finite number of distinct resources. It can be extended to a continuous distribution of resources whose use is determined by a multi-dimensional trait \citep{svardal_etal2014}. Here we use this extended model to study the long-term dynamics of multi-dimensional traits determining resource use. Besides the distribution of resources, there are two crucial quantities in these models. The first is a function that describes the rate of acquisition of different types of resources as a function of phenotype, and the second is a function that describes the rate of offspring production of a given phenotype as a function of the resource type consumed. In most models (in fact, in all previous models known to us), these functions are assumed to be essentially the same. Specifically, the rate of offspring production is simply proportional to the rate of consumption of a particular type of resource. However, biologically this is a rather special case, as it can easily be imagined that the type of resource that is most accessible (i.e., consumed at high rates) is not the type of resource that is most effectively converted into offspring. Here we show that with the traditional assumption of proportionality between the two functions, evolutionary dynamics of the multi-dimensional phenotype is always an equilibrium process. However, the complex dynamics of logistic competition models \citep{doebeli_ispolatov2014} can be recovered once this assumption is dropped. Thus, complicated dynamics are also prevalent in adaptive dynamics models that are based on explicit consumer-resource models. Our results show that it is crucial to carefully examine the assumptions made in such models, and in particular to make a clear a priori distinction between the rate of resource uptake and the rate at which a given resource is converted into offspring.

\section {Consumer-resource dynamics}
The consumer-resource system originally proposed
by MacArthur for a finite 
number of distinct resources \citep{macarthur_levins1967} can be generalized 
to a continuous spectrum of resources by assuming that different resource types are characterized by a generally multi-dimensional vector $z\in\mathbb{R}^n$ \citep{svardal_etal2014}.
Similarly, consumers are described by a multidimensional vector $x \in\mathbb{R}^{d}$. We let $R(z)$ denote the density of resource type $z$, and $C(x)$ denote the density of consumer type $x$. The rate at which resource $z$ is consumed by consumer $x$ is given by a function $\phi(x,z)$.  
The ecological dynamics of $R(z)$ is then described by a logistic equation with intrinsic
growth rate equal to 1 (without loss of generality) and a carrying capacity $K(z)$, plus a decay term describing resource consumption:
\begin{align}
\label{r1}
\frac{\partial R(z,t) }{\partial t}=R(z)\left( 1- \frac{R(z)}{K(z)}-\int \phi (x,z) C(x)dx\right).
\end{align} 
The integral in the consumption term reflects total consumption of resource $z$ by different consumer types $x$.

The dynamics of the consumer population, which is assumed to reproduce clonally, 
is determined by a birth term that incorporates the rate at which consumption of resource $z$ is converted into offspring of consumer type $x$. This rate is given by a function $\psi(x,z)$, which
is in general different from the consumption rate $\phi(x,z)$, reflecting the fact that the utility of a given resource is in general not solely determined by the rate of uptake of that resource. Note that all previous models of this this type are based on the assumption that $\psi(x,z)$ is proportional to $\phi(x,z)$.

Assuming a constant per capita death rate $\d$ for consumers, the ecological dynamics of consumer $x$ in the generalized MacArthur model is  
\begin{align}
\label{c1}
\frac{\partial C(x,t) }{\partial t}=C(x)\left(\int \psi (x,z) R(z)dz - \d \right).
\end{align} 
To analyze the evolutionary dynamics of the consumer type $x$, we make the assumption that the resource dynamics is much faster
than typical changes in the consumer density, an assumption that is routinely made in these kinds of models. Thus, the resource concentration is considered
to be completely relaxed to its steady state 
\begin{align}
\label{r2}
R^*[C(.),z]=K(z)\left( 1-\int \phi (x,z) C(x)dx\right),
\end{align} 
(where the notation $C(.)$ indicates that the resource steady state is a function of the current consumer distribution).
Then the consumer dynamics becomes
\begin{align}
\label{c2}
\frac{\partial C(x,t) }{\partial t}=C(x)\left[\int \psi (x,z) K(z)dz - \d - \int C(x')\int K(z)\psi (x,z)
\phi(x', z)dz dx'  \right].
\end{align} 
If we assume that the consumer is monomorphic for a given type $x$ (i.e., all consumer individuals have the same type $x$, which technically corresponds to assuming that the consumer distribution is a delta function centered at $x$), we obtain the equilibrium density of this monomorphic consumer, $C^*(x)$, by setting the right hand side of (4) to 0: 
\begin{align}
\label{c4}
 C^*(x)=\frac{\int K(z) \psi(x,z) dz - \d}{\int K(z) \psi(x,z) \phi(x,z) dz}.
\end{align}

\section{Adaptive dynamics}
Following the standard adaptive dynamics procedure \citep{geritz_etal1998}, we consider a monomorphic consumer resident $x$ and determine the invasion fitness $f(x,y)$ of a rare mutant consumer $y$. The invasion fitness $f(x,y)$ 
is the per capita growth rate of the rare mutant $y$ in the monomorphic resident population at its ecological equilibrium $C^*(x)$:
\begin{align}
\label{c3}
f(x,y)=\int \psi (y,z) K(z)dz -\d - C^*(x)\int  K(z)\psi (y,z)
\phi(x, z)dz,
\end{align} 
where $C^*(x)$ is given by (5).

Assuming for simplicity that the mutational variance-covariance matrix is the identity matrix \citep{leimar2009}, the adaptive dynamics of the trait $x$ is determined by the selection gradient $s(x)= \nabla_y f(y,x)|_{y=x}$:
\begin{align}
\frac{dx}{dt}=s(x),
\end{align}
where 
\begin{align}
\label{s1}
\nonumber s(x)&= \nabla_y f(y,x)|_{y=x}\\
&=\left[\int K(z) \psi(x,z) dz - \d \right] \times\\
\nonumber
&\left\{ \nabla_x \ln  \left| \int K(z) \psi (x,z) d z - \d \right|
-  \left. \nabla_y \ln  \left[ \int K(z) \psi (y,z)
  \phi(x,z) d z \right ] \right|_{y=x} \right\}.
\end{align} 
It is important to note that despite the name, the selection gradient is in general not a gradient of some function $h:\mathbb{R}^d\rightarrow\mathbb{R}$ on phenotype space (the selection gradient is only a gradient with respect to the mutant trait $y$, evaluated at the resident $x$). It is possible to give precise conditions in terms of the invasion fitness functions $f(x,y)$ for the selection gradient $s(x)= \nabla_y f(y,x)|_{y=x}$ to be the gradient of some function on phenotype space \citep{doebeli_ispolatov2013}. Here we can see directly that the selection gradient is a gradient of a function if we assume  that  the resource utilization rate $\psi (x,z)$ and uptake
rate $\phi(x,z)$ functions are functionally
related through a power law:
\begin{align}
\phi(x,z) = A \psi^{\a} (x,z). 
\end{align}
This includes of course the case where the two rates are proportional. Under this power law assumption, the selection gradient is reduced to an actual gradient form:
\begin{align}
\label{s2}
s(x)&=\left[\int K(z) \psi(x,z) dz - \d \right] \times\\
\nonumber
&\nabla_x \left\{ \ln  \left| \int K(z) \psi (x,z) dz - \d \right|
-  \frac{1}{\a+1}\ln  \left[ \int K(z) \psi^{\a+1} (x,z)]  dz \right ] \right\}.
\end{align} 
Because the selection gradient is in fact a gradient in phenotype space under these assumptions, the evolutionary dynamics (7) are a simple hill-coming process which equilibrates when the phenotype $x$ of the evolving consumer $x$ reaches the maximum of the function 
\begin{align}
\label{f1}
\frac {\left [\int K(z) \psi (x,z) dz - \d \right]^{\a+1}}
{ \int K(z) \psi^{\a+1} (x,z)  dz }.
\end{align} 

In particular, with the traditional assumption that the rates $\psi$ and $\phi$ are proportional, the single species evolutionary dynamics generated by the consumer-resource model are always equilibrium dynamics. We note that after reaching this equilibrium, adaptive diversification in the form of evolutionary branching into coexisting phenotypic species may or may not occur \citep{ackermann_doebeli2004, svardal_etal2014, geritz_etal2016}. In fact, such diversification becomes more likely with increasing dimension of phenotype space \citep{doebeli_ispolatov2014, svardal_etal2014,debarre_etal2014}. However, here we are not focussing on the problem of diversification, but rather on the evolutionary dynamics of single species, which always converge to an equilibrium under the above assumptions.

In fact, the assumption of a power law relationship between the uptake and utilization rates is also necessary for generating an evolutionary hill-climbing process. This can be seen by examining the second gradient on the right hand side of (8). For this gradient in $y$, evaluated at $y=x$, to be a gradient in $x$, the following relationship must hold
\begin{align}
\frac{d}{dx}\left\{\psi(x,z)\phi(x,z)\right\}=c\frac{\partial \psi(x,z)}{\partial x}\phi(x,z),
\end{align}
where $c$ is some constant. Dividing by $\psi(x,z)\phi(x,z)$ yields $\frac{\partial
  \{\psi(x,z) \phi(x,z)\}}{\partial x}/[\psi(x,z) \phi(x,z)]=c\frac{\partial \psi(x,z)}{\partial
  x}/\psi(x,z)$, and integrating both sides then yields
$\ln[\psi(x,z) \phi(x,z)]+D=c\ln(\psi(x,z))$ for some constant of integration
$D$, and hence the power law relationship between $\phi$ and $\psi$.

\section{Multiple resources}

It is worth noting that the above conclusions remain valid if a number of resource distributions $R_r(z_r)$ are available for consumption, as long as the total consumption is a linear combination of the consumption on each separate resource distribution. For example, if resource utilization is simply additive, the selection gradient generated in the consumer trait $x$ as a result of consuming multiple resources becomes, in analogy to eq. (8):
\begin{align}
\label{s1_many}
&s(x)=\left[\sum_r\int K_r(z_r) \psi_r(x,z_r) d z_r - \d \right] \times\\
\nonumber
&\left\{ \nabla_x \ln  \left| \sum_r\int K_r(z_r) \psi_r (x,z_r) d z_r - \d \right|
-  \left. \nabla_y \ln  \left[\sum_r \int K_r(z_r) \psi_r (y,z_r)
  \phi_r(x,z_r) d z_r \right ] \right|_{y=x} \right\}
\end{align} 
where $r$ indicates the different resource distributions $R_r(z_r)$, $z_r\in\mathbb{R}^{n_r}$. 
It is easy to see that the selection gradient (13) is a gradient in the phenotype $x$ if 1) there is a power law relationship $\phi_r(x,z_r)\propto\psi_r(x,z_r)^{\alpha_r}$ between the uptake rates $\phi_r(x,z_r)$ and the utilization rates $\psi_r(x,z_r)$ for each resource $R_r(z_r)$, and 2) the scaling exponent $\alpha_r=\alpha$ is the same for all $r$. The second condition is needed so that the second term on the right hand side of the single-resource gradient (10) generalizes to multiple resources (i.e., the condition is needed for 
the factor $1/(\alpha+1)$ in the second term on the right hand side of the single-resource gradient (10) to be the same for all resources). In particular, if the traditional assumption of $\psi_r(x,z_r)\propto\phi_r(x,z_r)$ holds for each resource distribution $R_r(z_r)$, then the resulting adaptive dynamics always converge to an equilibrium (which may or may not be evolutionarily stable).

\section{Complicated evolutionary dynamics}
If the two utilization and uptake rate functions $\psi (x,z)$ and $\phi(x,z)$ are not constrained by a power law relation, the selection gradient (\ref{s1})  can give rise to much richer class of evolutionary dynamics. To see this, we will show below that the selection gradient can then take the general form used in \cite{doebeli_ispolatov2014}, giving rise to adaptive dynamics of the form:
\begin{align}
\label{s3}
\frac{dx_i}{dt}=\sum_{j=1}^{d} b_{ij} x_j - x_i^3, 
\end{align} 
where the $x_i$ are the components of the phenotype vector $x$. In \cite{doebeli_ispolatov2014}, we have shown that such adaptive dynamics are likely to be chaotic when the dimension $d$ of phenotype space is large.

Assuming that the dimension of phenotype space $d$ is larger than the dimension of resource space $n$ (i.e., than the dimension of the vector $z$ in (1)), the selection gradient on the right hand side of (14) can be generated in the consumer-resource model with the following choices of $\psi (y,z)$ and $\phi(x,z)$:
\begin{align}
\label{s4}
\nonumber
\psi (x,z)&=\frac{1}{\sqrt{(2 \pi \Pi_{i=1}^{d}
    x_i^2)}}\exp\left( - \sum_{i=1}^{d}\frac{z_i^2}{2
    x_i^2}\right)F(z_{d+1},\ldots, z_{n})\\
\phi(x, z) &= \exp \left( \sum_{i=1}^{d}  z_i \sqrt{x_i^2- x_i^{-1}\sum_{j=1}^{d} b_{ij} x_j}
 \right) G(z_{d+1},\ldots, z_{n})/(\d-1)\\
\nonumber
K(z)&=\d+1.
\end{align} 
(Note that the expression for $\phi$ is only defined if the argument under the square root is $>0$.) The two functions $F$ and $G$ in $d-n$ variables are only restricted
by the normalization conditions
\begin{align}
\label{norm}
\int F(z_{d+1},\ldots, z_{n})  d
z_{d+1},\ldots, d z_{n}&=1,\\
\nonumber
\int F(z_{d+1},\ldots, z_{n}) G(z_{d+1},\ldots, z_{n}) d
z_{d+1},\ldots, d z_{n}&=1.
\end{align} 
Since the function $\psi$ is normalized,
\begin{align}
\label{mult}
\int K(z) \psi (x,z) d z=\d+1,
\end{align} 
the common multiplier in the selection gradient (8) becomes one, and the first gradient term ($\nabla_x\ldots$) on the right hand side of (8) is zero. 
The argument of the second gradient term ($\nabla_y\ldots$) on the right hand side of (8) is a Gaussian integral,
\begin{align}
\label{2term}
\int K(z) \psi(x,z) \phi(x,z) dz=\exp\left[ \sum_{i=1}^d y_i^2\left(x_i^2-\sum_{j=1}^db_{ij}x_j/x_i\right) /2  \right].
\end{align} 
Taking the logarithm and the gradient with respect to $y$ and evaluating at the resident $x$ yields 
the right hand side of (\ref{s3}) for
the $i$th component of the selection gradient.

\section{Conclusions}

Logistic competition models have long been used to study evolution of
traits influencing competitive interactions
\citep{roughgarden1979}. However, such models are usually based on
phenomenologically plausible assumptions about the functional form of
interactions between different phenotypes, rather than on a
mechanistic derivation of these functional forms following from how different phenotypes affect the ecological dynamics of a common resource. In fact, it has been noticed that the usual assumption of Gaussian forms for the functions describing interactions in logistic competition models is in general not compatible with explicit consumer-resource models \citep{taper_case1985, abrams1986, ackermann_doebeli2004}. In view of recent results regarding the occurrence of complex evolutionary dynamics in logistic competition models \citep{doebeli_ispolatov2014, regocosta_etal2017}, it is therefore interesting to see whether such dynamics can also be observed in consumer-resource models. 

In such models, there is a range of resources that are used
differently by distinct phenotypes. Here we considered generalized
MacArthur models \citep{macarthur_levins1967} in which both resource
distributions and consumer phenotypes are multivariate. In principle,
resource use by a consumer consists of two different biological
processes: resource uptake, and resource utilization, i.e., conversion of resources into offspring. In the consumer-resource models, these processes occur at rates $\phi(x,z)$ and $\psi(x,z)$, where $x$ is the consumer phenotype and $z$ is the resource type (both of which are generally multi-dimensional). The traditional assumption is that $\psi(x,z)\propto\phi(x,z)$, i.e., that utilization of a given resource type $z$ by a consumer of phenotype $x$ is proportional to the uptake of resource type $z$ by consumer type $x$, with the constant of proportionality independent of $x$ and $z$. However, biologically it may well be that a particular resource is easier to take up but less useful for producing energy (and offspring) than another type of resource that may be harder to acquire but may be more nutritious. Therefore, it is biologically reasonable to consider general consumer-resource models in which the uptake and utilization rates are not proportional.  

Using the framework of adaptive dynamics, we found that the assumption of proportionality between uptake and utilization rates places severe restrictions on the type of evolutionary dynamics that are generated by competition for resources. In fact, assuming that these rates are related by a power law, $\phi(x,z)\propto\psi(x,z)^\alpha$, the evolutionary dynamics are always a simple gradient process and therefore converge to an equilibrium of the adaptive dynamics. However, if these rates are not related through a power law, then we can recover the complicated evolutionary dynamics reported in \cite{doebeli_ispolatov2014}, who showed that chaotic evolutionary dynamics are common if the dimension of phenotype space is large. A general biological interpretation of these results is as follows. In models with explicit resource dynamics, the ecological interaction between different consumer types is mediated by consumption of the resource and conversion of the resource into offspring, and the mechanisms underlying the ecological interactions are subsumed in the consumption rates $\phi$ and the utilization rates $\psi$. For the resulting evolutionary dynamics of consumer types to be complicated, the ecological interactions mediated by resource use need to be complicated enough, which requires that the functions $\phi$ and $\psi$ are sufficiently different from each other, so that the relationship between consumption and utilization is sufficiently complicated. Here we have only provided a proof of principle of these ideas, and we have not attempted to give a specific biological interpretation of the type of rate functions needed to produce complex evolutionary dynamics. Nevertheless, our results show that it is important in principle to make the biological distinction between uptake and utilization rates in consumer resource models. While the traditional assumption that these rates are proportional always generates equilibrium dynamics, complex evolutionary dynamics can occur in consumer-resource models if the difference between uptake and utilization rates is pronounced enough.

\vskip 2cm

\noindent{\bf Acknowledgments:}  I. I. was supported by FONDECYT (Chile) grant 1151524. M. D. was supported by NSERC (Canada). 


\newpage
\bibliography{EvolutionofDiversity}
\bibliographystyle{evolution}

\end{document}